\documentclass[twocolumn]{cinc}
\pdfoutput=1
\usepackage{graphicx}
\usepackage{xcolor}
\usepackage{amsmath}
\usepackage{amssymb}
\usepackage{caption}
\usepackage{threeparttable}
\usepackage{multirow}
\usepackage{booktabs}
\usepackage{url}
\newcommand{\XZ}[1]{\textcolor{black}{#1}}

\newcommand{\Merlin}[1]{\textcolor{black}{#1}}

\usepackage{tcolorbox}

\begin{document}
\bibliographystyle{cinc}

\title{CardioRAG: A Retrieval-Augmented Generation Framework \\
for Multimodal Chagas Disease Detection}


\author {Zhengyang Shen$^{1}$, Xuehao Zhai$^{2}$, Hua Tu$^{1}$, Mayue Shi$^{1,3}$  \\
\ \\ 
$^1$ Department of Electrical and Electronic Engineering, Imperial College London, London, UK \\
$^2$ Department of Civil and Environmental Engineering, Imperial College London, London, UK \\
$^3$ Institute of Biomedical Engineering, Department of Engineering Science, University of Oxford, Oxford, UK }

\maketitle

\begin{abstract}
Chagas disease affects nearly 6 million people worldwide, with Chagas cardiomyopathy representing its most severe complication. In regions where serological testing capacity is limited, AI-enhanced electrocardiogram (ECG) screening provides a critical diagnostic alternative. However, existing machine learning approaches face challenges such as limited accuracy, reliance on large labeled datasets, and more importantly, weak integration with evidence-based clinical diagnostic indicators.

We propose a retrieval-augmented generation framework, CardioRAG, integrating large language models with interpretable ECG-based clinical features, including right bundle branch block, left anterior fascicular block, and heart rate variability metrics. The framework uses variational autoencoder-learned representations for semantic case retrieval, providing contextual cases to guide clinical reasoning. Evaluation demonstrated high recall performance of 89.80\%, with a maximum F1 score of 0.68 for effective identification of positive cases requiring prioritized serological testing. CardioRAG provides an interpretable, clinical evidence-based approach particularly valuable for resource-limited settings, demonstrating a pathway for embedding clinical indicators into trustworthy medical AI systems.

\end{abstract}

\begin{figure*}[!t]
    \centering
    \includegraphics[width=\linewidth]{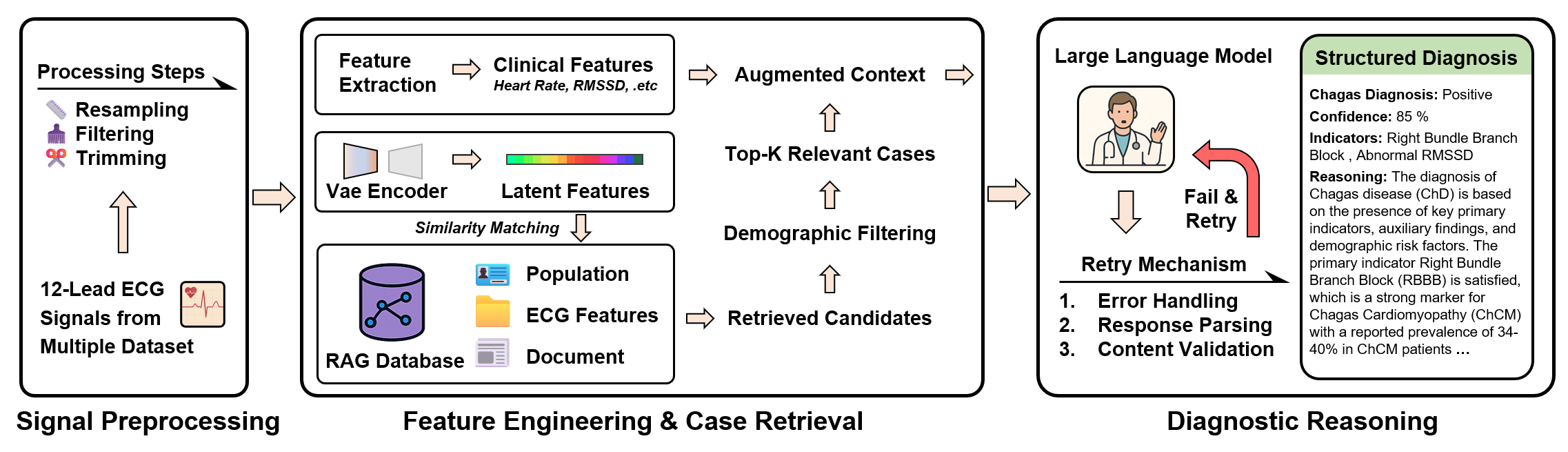}
    \vspace{-6mm}
    \caption{The CardioRAG Framework for Chagas disease diagnosis from 12-lead ECG signals. The system preprocesses raw ECG data, extracts clinical and latent features via VAE, retrieves relevant cases from a RAG database, and generates structured diagnoses with confidence scores using a large language model.}
    \label{fig:CardioRAGframework}
    \vspace{-2mm}
\end{figure*}

\section{Introduction}

Chagas disease is a neglected tropical disease caused by Trypanosoma cruzi, \XZ{affecting} approximately 6 million people worldwide, with fewer than 10\% aware of their infection status \cite{who_chagas_2025ChagasDisease}. The disease can progress to Chagas cardiomyopathy (ChCM), where electrocardiographic abnormalities often precede overt structural heart disease \cite{acquatella2007echocardiographyEchocardiographyinChagasheartdisease}. ECG provides a pragmatic, low-cost tool for early risk stratification in resource-limited settings, enabling prioritized serological testing and more efficient resource allocation \cite{alkmin2019Braziliannationalservice}.
This work addresses the 2025 PhysioNet Challenge focused on Chagas disease detection from ECG \cite{2025Challenge}.

\XZ{In recent years,} modern data-driven approaches have enabled new paradigms for disease detection from physiological signals. Advanced machine-learning methods can model non-linear relationships between disease status and multivariate time-series signals, including ECG \cite{silva2021prediction, jidling2023screening} and wearable sensor \cite{shen2025cobraIUPESM}.
However, current methods exhibit persistent limitations: (i) performance instability across domains due to population shift and limited calibration \cite{patrini2017Makingdeepneuralnetworks}, (ii) limited clinical interpretability hindering trust and adoption \cite{abbasian2024Interpretationofartificialintelligence}, and (iii) dependence on large, well-curated labeled datasets that are scarce for neglected diseases.

To address these challenges, we introduce CardioRAG, a novel multimodal retrieval-augmented generation (RAG) framework integrating interpretable ECG clinical features with large language model-based diagnostic reasoning. Our approach targets the critical screening scenario where high recall is essential for identifying potential Chagas cases for prioritized serological testing.

This work makes three key contributions: (1) A clinically-grounded RAG pipeline combining established ECG biomarkers (RBBB, LAFB) with heart rate variability metrics, achieving consistent high recall performance ($>$85\%) across different model configurations. (2) A VAE-based representation learning system coupled with demographic-aware case retrieval, enabling effective similarity matching with limited training data. (3) Empirical demonstration that prompt simplification and balanced case retrieval optimize performance for smaller language models, achieving 58.59\% accuracy and 87.76\% recall in zero-shot learning.

\section{Methodology}

We propose a comprehensive framework for automated Chagas disease detection that integrates deep learning-based ECG representation learning with retrieval-augmented generation (RAG) \cite{lewis2020retrieval} for enhanced diagnostic reasoning. As shown in Figure~\ref{fig:CardioRAGframework}, the system processes 12-lead ECG alongside patient demographic data (age, sex) recordings through three main stages: (1) extraction of clinical features from ECG signals, (2) VAE-based representation learning for semantic similarity \cite{kingma2014auto}, and (3) RAG-enhanced diagnostic decision making with large language models (LLMs).

\subsection{Data sources and preprocessing}
This study utilized three publicly available ECG datasets from the PhysioNet Challenge\cite{physionetchallengestandard}. The SaMi-Trop dataset \cite{cardoso2016longitudinal} (1,631 records, 400 Hz) contains validated positive cases from Brazil with serologically confirmed Chagas disease. The PTB-XL dataset \cite{wagner2020ptb} (21,799 records, 500 Hz) serves as negative controls from European patients in non-endemic regions. The CODE-15\% dataset \cite{ribeiro2020automatic} (300,000+ records, 400 Hz) provides mixed labels from Brazilian patients with self-reported Chagas status.

All ECG signals underwent standardized pre-processing: (1) resampling recordings to 400 Hz using linear interpolation, (2) standardizing signal durations to 7 seconds through cropping or padding, and (3) filtering using the NeuroKit2 toolbox \cite{makowski2021neurokit2} for noise removal and baseline correction.

\subsection{Chagas-specific feature engineering}

Chagas disease and ChCM manifest as specific ECG abnormalities, particularly conduction and rhythm disorders \cite{acquatella2007echocardiographyEchocardiographyinChagasheartdisease}. For conduction disorders, we implemented automated detection of right bundle branch block (RBBB) and left anterior fascicular block (LAFB) using Minnesota Code criteria \cite{prineas2010TheMinnesotacodemanualofelectrocardiographicfindings}. RBBB and LAFB represent key ChCM manifestations, with prevalence rates of 34-40\% and 23-39\% respectively in ChCM patients \cite{acquatella2007echocardiographyEchocardiographyinChagasheartdisease}. Table \ref{tab:ecg_Conduction_Disorders} outlines the specific ECG parameters required for automated detection.

\begin{table}[htbp]
\centering
\begin{threeparttable}
\caption{ECG parameters of conduction disorders}
\label{tab:ecg_Conduction_Disorders}
\small
\begin{tabular}{p{0.8cm}p{1.6cm}p{4.5cm}}
\hline
\textbf{Feature} & \textbf{Target Leads} & \textbf{Required ECG Parameters} \\
\hline
\multirow{4}{*}{RBBB} & \multirow{3}{*}{I, II, III, aVL} \multirow{3}{*}{aVF, V1, V2} & 
QRS duration, R wave duration, R peak duration, R wave amplitude, R' wave amplitude, S wave duration, S wave amplitude, net QRS deflection \\
\hline
\multirow{2}{*}{LAFB} & \multirow{1.2}{*}{I, II, III, aVL} \multirow{1.2}{*}{aVF} & 
QRS duration, Q wave duration, Q wave amplitude, QRS axis angle \\
\hline
\end{tabular}
\end{threeparttable}
\vspace{-4mm}
\end{table}

For rhythm assessment, RR-derived metrics were extracted from lead V5, including ventricular rate and RMSSD (root mean square of successive differences). RMSSD serves as a short-term heart rate variability index, with reduced values significantly associated with Chagas disease \cite{ribeiro2002Power-lawbehaviorofheartratevariabilityinChagas}. These features, combined with demographic information (age and sex), form the comprehensive multimodal input to the RAG diagnostic system.

\subsection{CardioRAG diagnostic architecture}

The RAG framework addresses the fundamental challenge of labeled data scarcity in Chagas disease detection by enabling case-based reasoning \XZ{via} retrieval of similar historical cases. This diagnostic approach aligns with clinical practice, in which physicians rely on prior cases to guide complex diagnostic decisions. \cite{ng2025RAGinhealthcare, lewis2020retrieval}.

\textbf{Variational autoencoder for signal embedding.} We employ a variational autoencoder (VAE) architecture \cite{kingma2014auto} \XZ{to learn} compact ECG representations that \XZ{support effective similarity search}. The encoder consists of four residual blocks with progressively increasing channels (32, 64, 128, 256). \XZ{Each residual block contains two 1D convolutions with Batch Normalization, ReLU and a skip connection.} \XZ{The encoder outputs} ($\mu$) and log-variance ($\log\sigma^2$) parameters of a 256-dimensional latent distribution. Training employs the standard VAE objective:
\begin{equation}
{L} = {L}_{\text{recon}} + \beta \cdot {L}_{\text{KL}}
\end{equation}
where ${L}_{\text{recon}} = {E}_{q_\phi(z|x)}[\log p_\theta(x|z)]$ is the reconstruction loss, ${L}_{\text{KL}} = D_{\text{KL}}(q_\phi(z|x)||p(z))$ is the KL divergence regularization term, and \XZ{$\beta$ is set to $0.1$ based on validation performance.}

\textbf{Case retrieval mechanism.} The retrieval process implements a two-stage search strategy combining VAE-based similarity with demographic filtering. Similarity search begins in the VAE latent space using cosine similarity to identify the k most similar cases \XZ{(with k tuned on validation data)}. The secondary filtering computes a composite similarity score:
\begin{equation}
S_{\text{composite}} = S_{\text{VAE}} + w_{\text{age}} \cdot S_{\text{age}}
\end{equation}
where $S_{\text{VAE}}$ is normalized VAE similarity, $S_{\text{age}}$ reflects age similarity using a Gaussian kernel with $\sigma = 10$ years, and $w_{\text{age}}$ is the weighting coefficient. Retrieved cases are formatted into structured context for the LLM, including patient demographics, detected clinical features, HRV metrics, and diagnostic labels\XZ{, with length control to avoid prompt overflow}.

\textbf{LLM powered diagnostic reasoning.} The LLM receives structured prompts containing patient features and retrieved similar cases, generating diagnostic predictions with confidence scores and clinical reasoning. 
The LLM output follows a structured JSON format containing: (1) binary diagnosis (POSITIVE/NEGATIVE), (2) confidence percentage, (3) detailed clinical reasoning, (4) identified diagnostic indicators, (5) relevant risk factors, and (6) other cardiac findings. Note while confidence scores are generated, we found them unreliable for smaller LLMs and thus focus evaluation on binary diagnostic performance.



\begin{tcolorbox}[colback=gray!10, colframe=gray!50, title=LLM-generated diagnostic rationale]
\small
The patient presents with RBBB\_satisfaction indicating a right bundle branch block, which is consistent with Chagas. The low RMSSD in Lead V6 (7.8 ms) strongly suggests a heart rhythm abnormality indicative of Chagas. No other significant ECG findings are noted, and the data supports a clear positive diagnosis.  (Chagas positive)
\small
\end{tcolorbox}

\section{Results and analysis}
\vspace{-1mm}

Our CardioRAG framework could not be evaluated using the standard PhysioNet Challenge 2025 methodology due to technical constraints: the local storage limit prohibits inclusion of large language models or API connectivity required for our system. Additionally, our zero-shot learning paradigm fundamentally differs from the Challenge's supervised training approach.

Therefore, we evaluated the proposed framework using the DeepSeek-R1:1.5b language model \cite{deepseekai2025deepseekr1incentivizingreasoningcapability} on a test set of 100 patients, consisting of 50 consecutive positive cases from the SaMi-Trop dataset and 50 consecutive negative controls from the PTB-XL dataset. Our experiments focused on two critical aspects: the impact of prompt engineering, and the effect of RAG retrieval strategies on diagnostic performance.

\subsection{Prompt engineering}

\vspace{-6mm}
\begin{figure}[!h]
    \centering
    \includegraphics[width=0.4\textwidth]{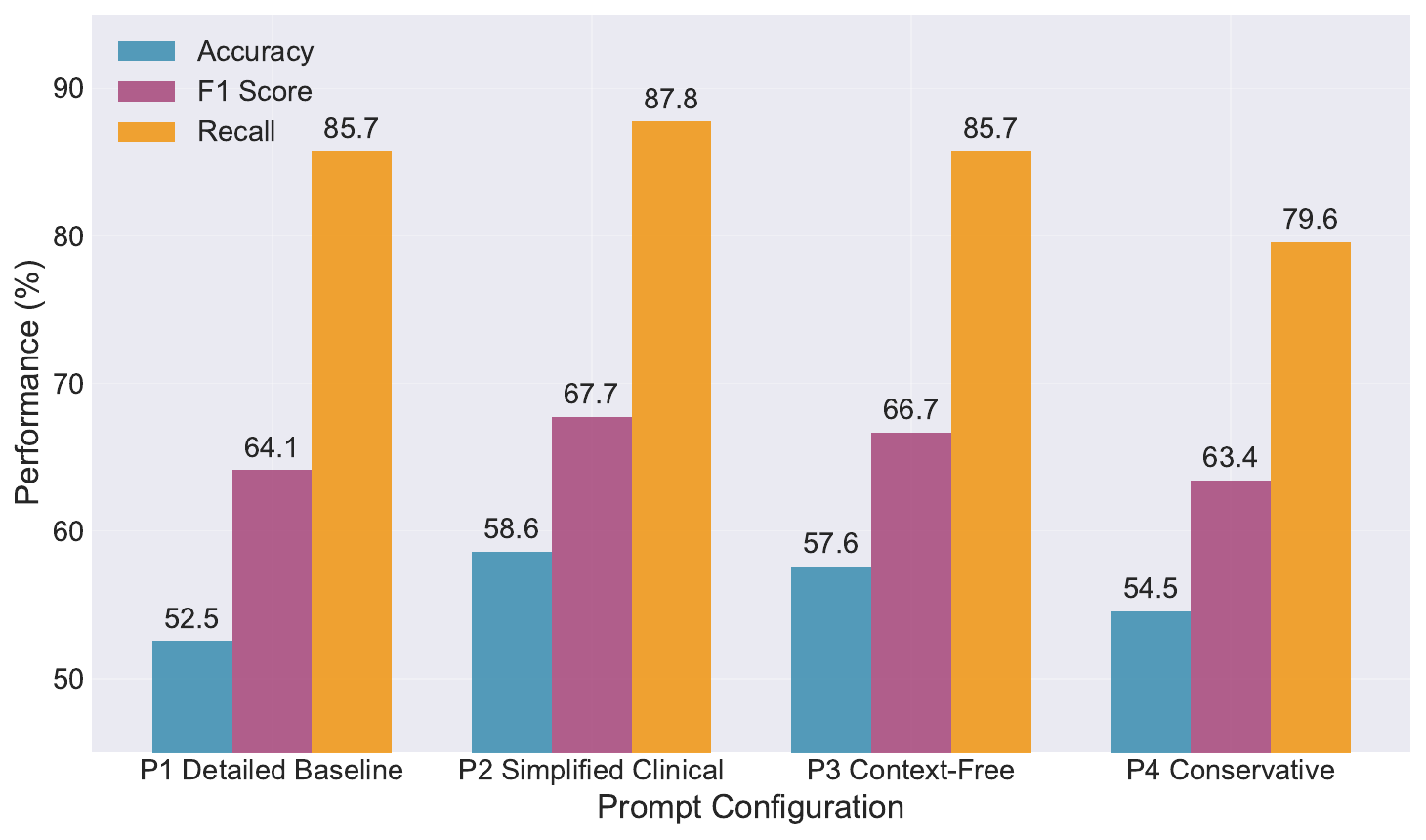}
    \vspace{-4mm}
    \caption{Impact of prompt engineering (top-k retrieved case, k=8). Configurations: P1 Detailed prompt (baseline, full ECG criteria and clinical instructions), P2 Simplified Clinical (without detailed ECG criteria for RBBB/LAFB), P3 Context-Free (without diagnostic background), P4 Conservative (includes cautionary guidance for positive diagnoses).}
    \label{fig:prompt_engineering}
    \vspace{-4mm}
\end{figure}

Figure~\ref{fig:prompt_engineering} presents the performance comparison across four prompt configurations. Counterintuitively, the "P2 Simplified Clinical" configuration achieved the best performance with 58.59\% accuracy, 87.76\% recall, and 67.72\% F1 score, representing significant improvements over the "P1 Detailed Baseline" (52.53\% accuracy, 85.71\% recall, 64.12\% F1). This 6.06 percentage point accuracy improvement suggests that for smaller language models, concise prompts focusing on key decision factors outperform exhaustive clinical descriptions with detailed RBBB/LAFB detection criteria.

Notably, adding cautionary instructions ("P4 Conservative") decreased performance to 54.55\% accuracy, indicating that overly conservative prompting may bias the model toward indecision. The optimal configuration maintained essential clinical context while avoiding information overload. In the annalysis, one case could not produce a valid structured output from the language model and was therefore excluded from the subsequent evaluation.

\subsection{Retrieval strategies}
\vspace{-1mm}

Table~\ref{tab:results_summary} demonstrates the impact of retrieval augmentation on diagnostic performance. The relationship between the number of retrieved cases (k) and accuracy follows an inverted U-shape, with optimal performance at k=8 (58.59\% accuracy). The baseline prompt (P1) without RAG achieved a markedly low recall of 48.98\%, which is significantly lower than all configurations with RAG. This demonstrated that RAG effectively enhanced the LLM’s diagnostic performance.

\begin{table}[h]
\centering
\caption{\label{tab:results_summary}Comparison of retrieval configurations}
\vspace{-2 mm}
\small
\begin{tabular}{lccc} \hline\hline
Configuration              & Accuracy\% & Recall\% & F1 Score \\ \hline
P1 No RAG           & 54.55 & 48.98 & 0.52 \\
\Merlin{P1 RAG k=8}	         & 52.53 & 85.71 & 0.64 \\
P2 RAG k=8             & 58.59 & 87.76 & 0.68 \\
P2 RAG k=8 (bal)       & 58.59 & 89.80 & 0.68 \\
P2 RAG k=16            & 52.53 & 77.55 & 0.62 \\ \hline\hline
\hline
\end{tabular}
\vspace{-4 mm}
\end{table}


The performance degradation observed at k=16 (52.53\% accuracy) may be attributed to the introduction of excessive retrieved cases, which likely added noise rather than providing useful diagnostic context and potentially overwhelmed the LLM’s reasoning capacity. In contrast, the balanced retrieval strategy at k=8 achieved the highest recall and F1 score, suggesting the importance of maintaining an appropriate proportion of representative positive and negative examples within the retrieval set.

These findings suggest alignment with our prompt engineering results, indicating that both prompt quality and RAG quantity may significantly influence LLM diagnostic performance. Our results show that neither maximal information provision nor extreme simplification yields optimal performance. Instead, balanced, focused contextual guidance appears to achieve the best diagnostic reasoning outcomes without cognitive overload.

\section{Conclusion}
\vspace{-1 mm}

Our CardioRAG framework demonstrates the potential of integrating retrieval-augmented generation with clinical ECG features for Chagas disease screening, achieving 58.59\% accuracy and 87.76\% recall with consistently high recall across configurations. Our evaluation reveals that simplified prompts outperformed detailed descriptions; moderate case retrieval (k=8) with balanced retrieval achieved optimal performance; and the 58-59\% accuracy ceiling may reflect current model limitations, warranting evaluation of larger LLMs. The framework's high recall performance makes it valuable for initial screening and patient triaging for serological testing, with future work focusing on improving specificity through enhanced feature selection and RAG optimization.


\bibliography{refs}


  
  
      

\vspace{-2 mm}
\begin{correspondence}
Mayue Shi\\
Institute of Biomedical Engineering,
University of Oxford, Oxford OX3 7DQ, UK.\\
mayue.shi@eng.ox.ac.uk and m.shi16@imperial.ac.uk.
\end{correspondence}

\end{document}